# Characterization of carbon contamination under ion and hot atom bombardment in a tin-plasma extreme ultraviolet light source

***A Dolgov[1], D Lopaev[2], C J Lee[1], E Zoethout[4], V. Medvedev[1], O Yakushev[3] and F Bijkerk[1]**

[1] MESA+ Institute for Nanotechnology, University of Twente, Enschede, The Netherlands
[2] Skobeltsyn Institute of Nuclear Physics, Moscow State University, Moscow, Russian Federation
[3] Institute for Spectroscopy, Moscow, Russian Federation
[4] Dutch Institute for Fundamental Energy Research (DIFFER), Nieuwegein, The Netherlands

E-mail: a.dolgov@utwente.nl

**ABSTRACT**

Molecular contamination of a grazing incidence collector for extreme ultraviolet (EUV) lithography was experimentally studied. A carbon film was found to have grown under irradiation from a pulsed tin plasma discharge. Our studies show that the film is chemically inert and has characteristics that are typical for a hydrogenated amorphous carbon film. It was experimentally observed that the film consists of carbon (~70 at. %), oxygen (~20 at. %) and hydrogen (bound to oxygen and carbon), along with a few at. % of tin. Most of the oxygen and hydrogen are most likely present as OH groups, chemically bound to carbon, indicating an important role for adsorbed water during the film formation process. It was observed that the film is predominantly sp3 hybridized carbon, as is typical for diamond-like carbon. The Raman spectra of the film, under 514 and 264 nm excitation, are typical for hydrogenated diamond-like carbon. Additionally, the lower etch rate and higher energy threshold in chemical ion sputtering in $H_2$ plasma, compared to magnetron-sputtered carbon films, suggests that the film exhibits diamond-like carbon properties.

*Corresponding author. E-mail: dolgov.adonix@gmail.com; phone: +31622892158 (Netherlands)

## 1.INTRODUCTION

The phenomenon of the contamination of optics under the action of EUV radiation is a known problem that is actively studied. Even in vacuum conditions, carbon atoms are deposited onto a surface when residual hydrocarbons dissociate during interaction with energetic photons [1]. The carbon atoms can then accumulate on surfaces, which is undesirable for reflective optics because the carbon layer absorbs radiation. For instance, EUV-induced carbon contamination of grazing incidence optics in synchrotron beam lines is one of the major reasons for the optics' reduced reflectivity. The study of this problem began a few decades ago [2], and several cleaning mechanisms have since been investigated [ 3-5].

Extreme ultraviolet lithography (EUVL) is currently the most advanced technology for the fabrication of integrated circuits with characteristic half-pitch ≤ 22 nm. In modern 13.5 nm EUV lithography, the main contaminant on optics is carbon. Even at pressures as low as $10^{-5}$-$10^{-6}$ Pa, the C film deposition rate can be as high as 0.01–1 nm/h for an average EUV-radiation intensity of about 0.1-1 W/cm$^2$. A carbon film will cause a reflectivity loss of about 1%/nm per mirror, which is especially critical for optical systems that have multiple reflecting surfaces.

Off-line removal of EUV-induced carbon contamination also reduces the duty cycle of EUVL, which is undesirable [1, 6, 7], and ultimately, increases the operating cost of the EUVL process. However, with the use of fluxes of ions or neutral reactive species to restore reflectivity, the lifetime of a standard Mo/Si MLM system can be extended to several thousand hours [8]. A new cleaning strategy that uses the plasma, induced by EUV ionization of the low-pressure gas (usually $H_2$) over the mirror surface, was recently proposed [9]. This inline cleaning process has the potential to increase the duty cycle of EUVL.

Until recently, magnetron deposited carbon was frequently used as a model for EUV-induced carbon growth. However, the characteristics of a carbon film depend on the deposition conditions. In practice, this means that the optics located close to the EUV light source (i.e., collector mirrors) could be coated with a significantly denser carbon film, due to the presence of a broad spectrum of high energy photons and ions. The structure of EUV-induced carbon films determines the reflectivity losses of the mirror. For instance, the denser structure of diamond-like carbon (DLC) film leads to more EUV absorption compared to a polymeric layer or "soft" amorphous carbon layer of the same thickness. Furthermore, because the carbon in DLC films is more tightly bound, the films are likely to be more difficult to clean. Hence, it is necessary to know how the phase of carbon may vary depending on the growth conditions.

*Corresponding author. E-mail: dolgov.adonix@gmail.com; phone: +31622892158 (Netherlands)

In the presence of water and heavy hydrocarbons, a carbon film is likely to be hydrogenated. Typically, hydrogenated amorphous carbons are classified into four classes, illustrated by the phase diagram shown in Fig. 1: [10]
]

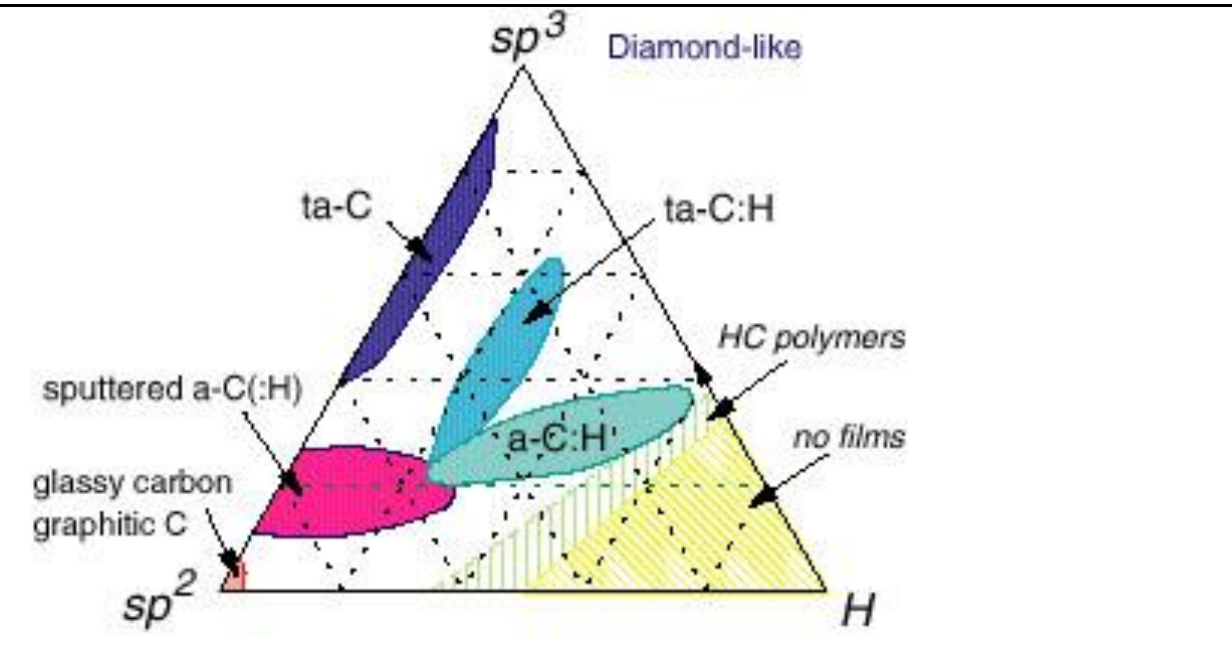

*Figure 1. Phase diagram of hydrogenated amorphous carbon films.*

All a-C:H films have a specific hydrogen content and properties that are revealed by applying a combination of different diagnostic techniques, such as X-ray photoelectron spectroscopy (XPS), Raman spectroscopy (RS), energy-dispersive X-ray (EDX) spectroscopic analysis, electron energy loss spectroscopy (EELS), transmission electron microscopy (TEM), and others.

In this paper, we report the experimental study of carbon-based films, deposited on the collector mirror of an EUV source. The condition under which the film was grown will be described, as well as analysis showing that the film has DLC characteristics.

## 2.EXPERIMENT

### *2.1.Film deposition*

The experimental set up is schematically shown in figure 2. A *Z*-pinch discharge in Sn vapour is used as a source of pulsed EUV emission. The discharge is initiated using an Nd:YAG laser pulse to evaporate Sn from the surface of a liquid tin cathode. The Sn vapour short circuits the 3 mm gap to the anode, which is biased at high voltage (3–4 kV). A few nH inductance of the discharge circuit allows a highly ionized Sn ($Sn^{8+-11+}$) plasma to be produced, which with further pinching leads to a powerful shot of EUV emission being generated. The energy input into the discharge is 1–2 J, while CE (conversion efficiency) to 13.5 nm emission in a 2% spectral band is about 1%. The shot repetition rate of the EUV source is 1.5 kHz, which allows a high photon fluence for a period of ~3 hours with 3–5% stability.

*Corresponding author. E-mail: dolgov.adonix@gmail.com; phone: +31622892158 (Netherlands)

The *Z*-pinch emission is collected by a set of grazing incidence cylindrical Mo mirrors, called the EUV collector. Our test bench collector consists of six mirrors, located at a distance of 48 cm from the source of radiation. Each mirror consists of a curved glass-ceramic sitall substrate that is 45 mm x 60 mm square, and has a thickness of 6 mm. High reflectivity EUV range is achieved by coating the substrate with 50 nm of Mo.

To protect the collector from the products of the Z-pinch discharge, a series of traps are located between the discharge and the mirrors. The debris mitigation system consists of a magnetic field to defect most ions with an energy less than 100 keV and rotating foil trap, which effectively collects all micron-sized tin droplets moving at speeds of less than 500 m/s. Thus, the collector is exposed to radiation (both in-band EUV, and deep ultraviolet), high energy ions, and high velocity Sn debris during source operation.

Before the experiments, the source chamber was evacuated to 3–5 × $10^{-6}$ Torr. The partial pressures of the background gases were: water ~$10^{-7}$ Torr, nitrogen ~$10^{-7}$ Torr oxygen ~$10^{-9}$ Torr, and residual hydrocarbons $10^{-9}$ Torr. The experimental set up is schematically shown in figure 2. During operation, $H_2$ is fed into the vacuum channel, where the collector mirrors are located. The hydrogen flows in the direction of the EUV source chamber, slowing down and scattering energetic ions and small Sn droplets in collisions with neutral and charged particles of $H_2$ plasma formed in the collector optics channel. The presence of $H_2$ over the collector mirrors notably decreases the contamination rate of the mirrors by carbon. The $H_2$ pressure in the channel is ~$10^{-3}$ Torr and does not influence on EUV source operation.

*Corresponding author. E-mail: dolgov.adonix@gmail.com; phone: +31622892158 (Netherlands)

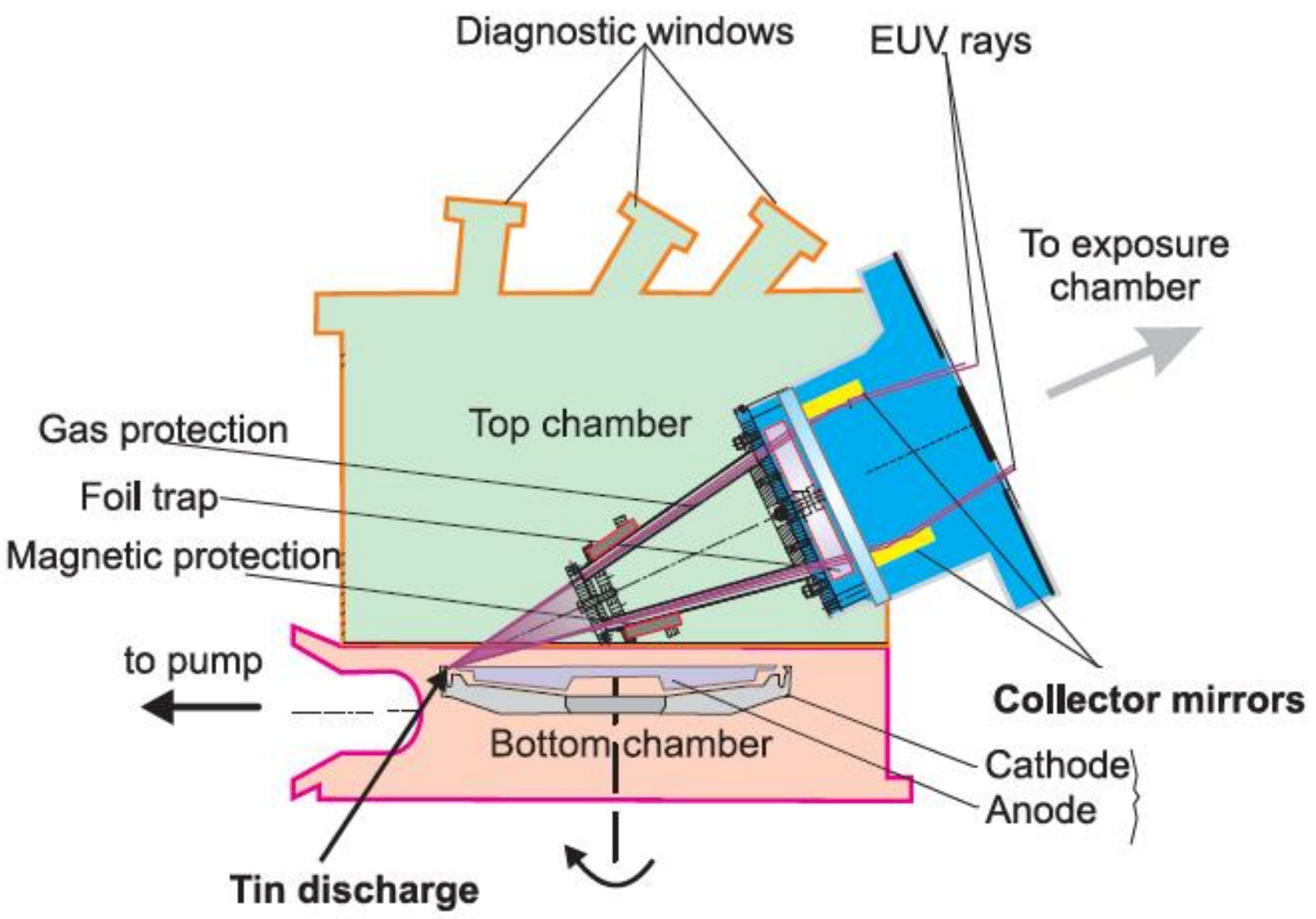

Figure 2. EUV tin plasma source for direct reproducing EUV-lithography tool conditions.

## 3.ANALYSIS

Our contamination film (see figure 3 OPTICAL image) was gray-brown in color, and transparent in the visible region. The films, which were still attached to the collector mirror, were placed in 65% $HNO_3$ acid. It was noticed that, even after the underlying molybdenum was etched away (~20 seconds) with nitric acid, the carbon film remained on the sital substrate. The undamaged carbon film flakes were then removed from the substrate and transferred to a copper support grid with a 1 X 1 mm mesh, and then sandwiched in a slotted copper envelop grid to provide mechanical stability.

The total thickness of the film was found to be of the order of 150 nm after few hundreds millions EUV shots. The thickness was measured by EELS, in the energy-filtered transmission electron microscopy (EFTEM) mode. At this thickness, a film of magnetron sputtered carbon would be expected to be completely etched away within 10 seconds in 65% $HNO_3$ acid.

*Corresponding author. E-mail: dolgov.adonix@gmail.com; phone: +31622892158 (Netherlands)

| *Table 1. Atomic compounds of collector grown carbon film according to energy-dispersive X-ray spectroscopic analysis (EDX)* | | | | | | | | | | | |
|---|---|---|---|---|---|---|---|---|---|---|---|
| *Area 1* | | | *Area 2* | | | *Area 3* | | | *Average* | | |
| *Element Line* | *Atom %* | *Atom % Error* | ***Element Line*** | ***Atom %*** | ***Atom % Error*** | *Element Line* | *Atom %* | *Atom % Error* | *Element Line* | *Atom %* | *Atom % Error* |
| ***C K*** | **72.91** | **+/- 1.57** | ***C K*** | **72.08** | **+/- 1.16** | ***C K*** | **70.36** | **+/- 1.39** | ***C K*** | **71.78** | **+/-1.37** |
| ***O K*** | **19.00** | **+/- 0.41** | ***O K*** | **18.35** | **+/- 0.37** | ***O K*** | **17.88** | **+/- 0.36** | ***O K*** | **18.41** | **+/-0.38** |
| ***Sn L*** | **2.62** | **+/- 0.07** | ***Sn L*** | **3.18** | **+/- 0.05** | ***Sn L*** | **3.39** | **+/- 0.07** | ***Sn L*** | **3.06** | **+/-0.06** |

To understand the atomic composition of the film, EDX was used. According to EDX results (see table 1) the film is, on average, composed of the order of approx. 72 at. % C, and approx. 18 at. % O. In addition, trace elements from the EUV source fuel (Sn), electrodes (Fe, Ni, and Mg), and the collector mirror (Mo, and Si) were also found. The sample was also slightly contaminated by the attempts to chemically etch it (Cr, P, and others). Despite these low levels of various elements, the film is predominantly carbon, oxygen and tin, which is expected. We assume that the carbon and oxygen content of the film comes from background gases and adsorbed water.

Within the film, there are larger particles that are denser than the surrounding film. EELS spectra (data not shown) for the film are typical for amorphous carbon. Secondary electron microscopy (SEM), TEM, and high resolution TEM (HRTEM) images show the presence of submicron tin droplets. The large droplets are not randomly distributed. The droplet tracks are parallel to each other, indicating that they are ballistic tin droplets that have travelled directly through the rotating foil trap (part of mitigation protection system) and have a velocity more than 500 m/s (the minimum speed to allow the particles to pass through the foil trap).

Interpretation of the HRTEM carbon lattice images gives fringes with d-spacings of 0.15 and 0.25 nm in the areas from 2 to 10 nm. Such lattice structure are similar to ion-beam deposited carbon and carbon produced by shock synthesis designated as "0.25-i-carbon" and presented in [11]. This amorphous carbon can be attributed to hard carbon coatings or DLC. Only in one case the lattice fringes yielding a d-spacing of 0.205 +/- 0.001 nm, associated with (111) cubic or n-diamond lattice plane, were found. However, no evidence for the 0.178 nm and 0.126 nm d-spacings, associated with the (002) and (022) diamond lattice planes, were found.

Atomic force microscopy (AFM) measurements indicate that the surface is relatively rough, with an RMS roughness of a few tens of nm.

*Corresponding author. E-mail: dolgov.adonix@gmail.com; phone: +31622892158 (Netherlands)

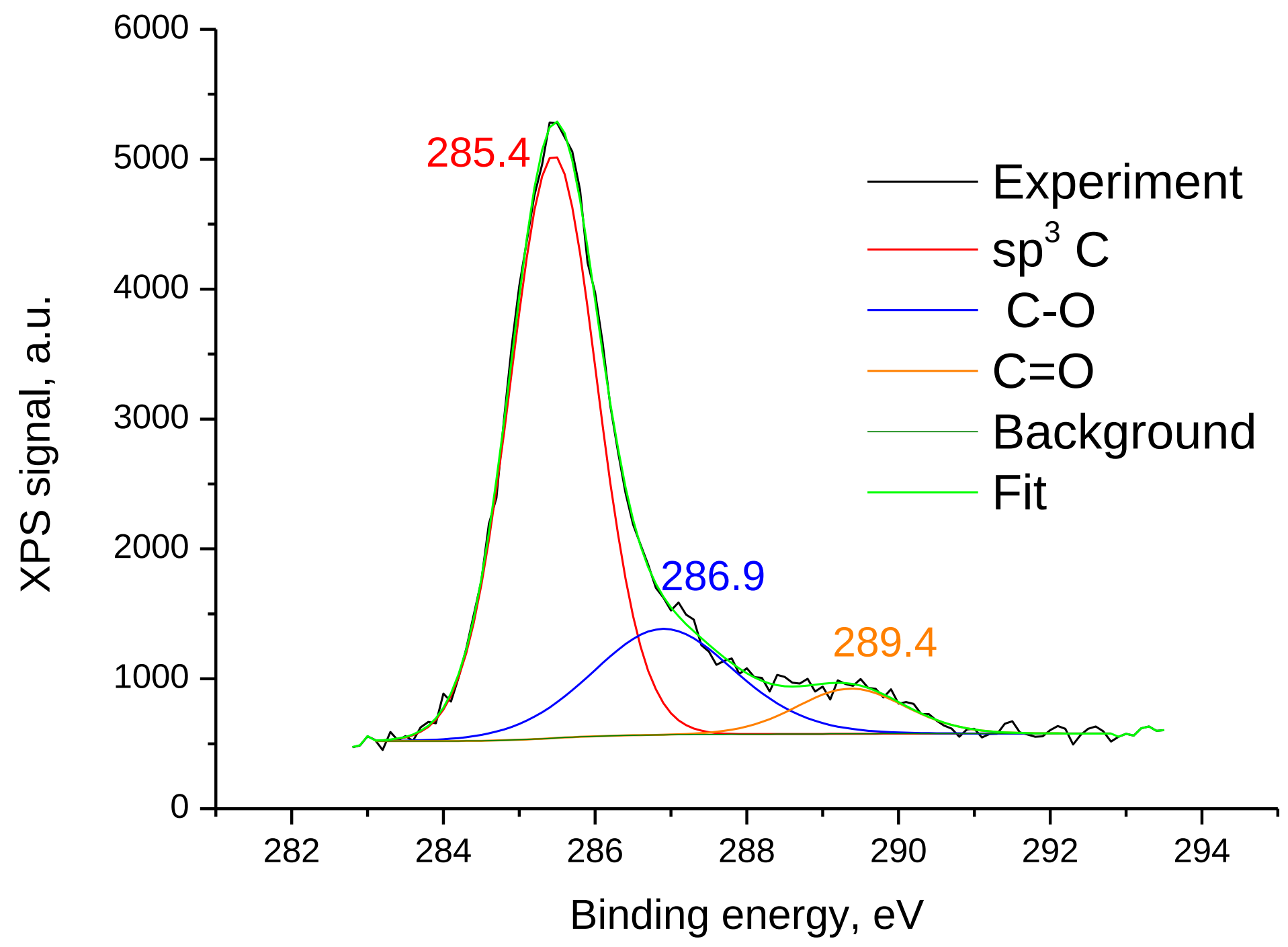

Figure 4. XPS results for carbon C1s peak. Three peaks at 285.4 eV , 286.9 eV and 289,4 eV are attributed to the $sp^3$ bulk bonded carbon C-C, C-O chemical bonds and C=O bonds.

More accurate information on the structural properties of carbon films was obtained with an XPS study (Thermo Theta Probe monochromated Al K**α)**. Figure 4 shows the characteristic C 1s peak from the carbon film. The atomic percent of carbon was determined to be 70±2%, in agreement with the EDX data (tab 1). Since XPS is a surface sensitive technique, whereas EDX probes the complete 150 nm thick carbon film, similar composition indicates a homogeneous in-depth composition.

Deconvolution was carried out after subtracting the background by Shirley's method. The peaks were deconvolved into Gaussian-Lorenzian mix functions. The deconvolved high resolution C 1s spectrum of film yielded three peaks at 285.4 eV , 286.9 eV and 289,4 eV, which are attributed to the $sp^3$ bulk bonded carbon C-C, C-O chemical bonds and C=O bonds (Fig. 4). The first peak at around 285.4 eV is associated with C-C $sp^3$ hybridization. This can be attributed to sp3 C-H bonding, or a diamond like structure. Peak that observed at 286.9 eV could be C-O-C or C-O-H. Another sub-peak was observed at around 289.4 eV, and is assigned to the C(O)OH or HO-C-OH bonds [12, 13] It should be noted that we **cannot** fit the data with an $sp^2$ peak, but the presence of $sp^3$ by itself does not confirm the DLC nature of the film, since it is possible that most sp3 bonds are terminated by hydrogen, which cannot be directly detected with the techniques used here.

*Corresponding author. E-mail: dolgov.adonix@gmail.com; phone: +31622892158 (Netherlands)

Raman spectroscopy was used to characterize the molecular content of the film. Wavelengths of 264 and 514 nm were used to excite the Raman spectra. Additionally, Raman maps of the film were collected with a Jobin Yvon T64000 spectrometer at 514 nm and home-built UV Raman imaging system, based on a Ti-sapphire laser (the signal was measured on 4th harmonic) system at 264 nm with a resolution of 1 $cm^{-1}$.

Measured Raman spectra of collector-grown carbon for ultraviolet (264 nm) and visible (514 nm) excitation wavelengths are presented in figure 5. All carbon films show similar features in their Raman spectra in the 700–2000 $cm^{-1}$ region. The spectrum for 514 nm excitation was deconvolved into two Gaussians with maxima in the regions of the so-called D- (Disordered), and G- (Graphite) peaks, around 1360 $cm^{-1}$, and 1580 $cm^{-1}$, respectively. For UV excitation, a peak at ~1060 $cm^{-1}$ due to C-C *sp*3 vibrations, called the T peak, is usually observed [14-16]. The location, width, and intensity of G-, D- and T-peaks in multi-wavelength Raman spectrum provide a unique signature, which we use to characterize the carbon structure.

For UV excitation, the G peak position was found to be centered at 1615 $cm^{-1}$, which is blue shifted compared to typical a-C (1580 $cm^{-1}$), or nanocrystalline graphite (1600 $cm^{-1}$). The G peak position and its dispersion (see fig 5 b) is very close to that found for ta-C:H [17], or an annealed ta-C:H film. The latter exhibits a strong D peak under 514 nm excitation, but shoulder-like D peak under 264 nm excitation [18]. The carbon film grown on the collector mirror most likely consists of tetrahedral amorphous carbon (ta-C).

The UV Raman spectrum exhibited no explicit T peak. The reason could be a presence of sufficient $sp^2$ content of ta-C (oxygen bonds), which normally reduces the *T* peak intensity. Instead, a D-like peak, centered at 1410 $cm^{-1}$, is observed. The D-like peak is associated with the breathing mode of ring structures, however, the ring's symmetry requires a nearby defect to excite the breathing mode. Thus, the presence of the D-like peak indicates that the film has a substantial fraction of carbon atom rings, but with a significant amount of disorder [14-18]. This feature in the UV Raman spectrum, in the absence of a T peak, is attributed to the C–C stretch of both hydrogenated and hydrogen free $sp^3$ sites. From this combination of factors, we conclude that the film consists of ta-C:H film

The presence of hydrogen in the film is confirmed by Raman spectroscopy. A broad pair of peaks, shown in Fig. 5b, in the range 2500-3200 $cm^{-1}$ cannot be attributed to the second order of the D peak, known as the 2D' peak. Instead, we attribute this intensive peak to two sources: sp3

*Corresponding author. E-mail: dolgov.adonix@gmail.com; phone: +31622892158 (Netherlands)

$CH_x$ stretching vibrations at 2960 $cm^{-1}$ and, the second, at ~3210 $cm^{-1}$, is due to the second order G- peak [19].

Since terminated C-H bonds have a large dipole moment they can be observed directly using IR spectroscopic techniques. We used reflection absorption infrared spectroscopy (RAIRS) to observe sp3 $CH_x$ bonds. The RAIRS spectrum of the collector film in the range 2800-3100 $cm^{-1}$ is shown in Figure 6. Although the absorbance of sp3 $CH_2$ and sp3 $CH_3$ bonds in RAIRS is small, they are, nevertheless, clearly visible in Figure 6.

The amount of hydrogen (bound to carbon) in the film can be estimated from the ratio of the slope, *m*, of the fitted linear (fluorescent) background, and the intensity of the *G* peak, $m/I(G)$ [19]. As shown in Fig. 5, the ratio m/$I$(G) was found to be 0.5 μm , which corresponds to ~20 at. % of hydrogen. Taking into account that the XPS data does not distinguish between C-C and C-H bonds, we can conclude that the film has ~20 % alcohol and aldehyde groups, ~20 % $CH_x$ groups, while the rest of it is $sp^3$ C-C bonds.

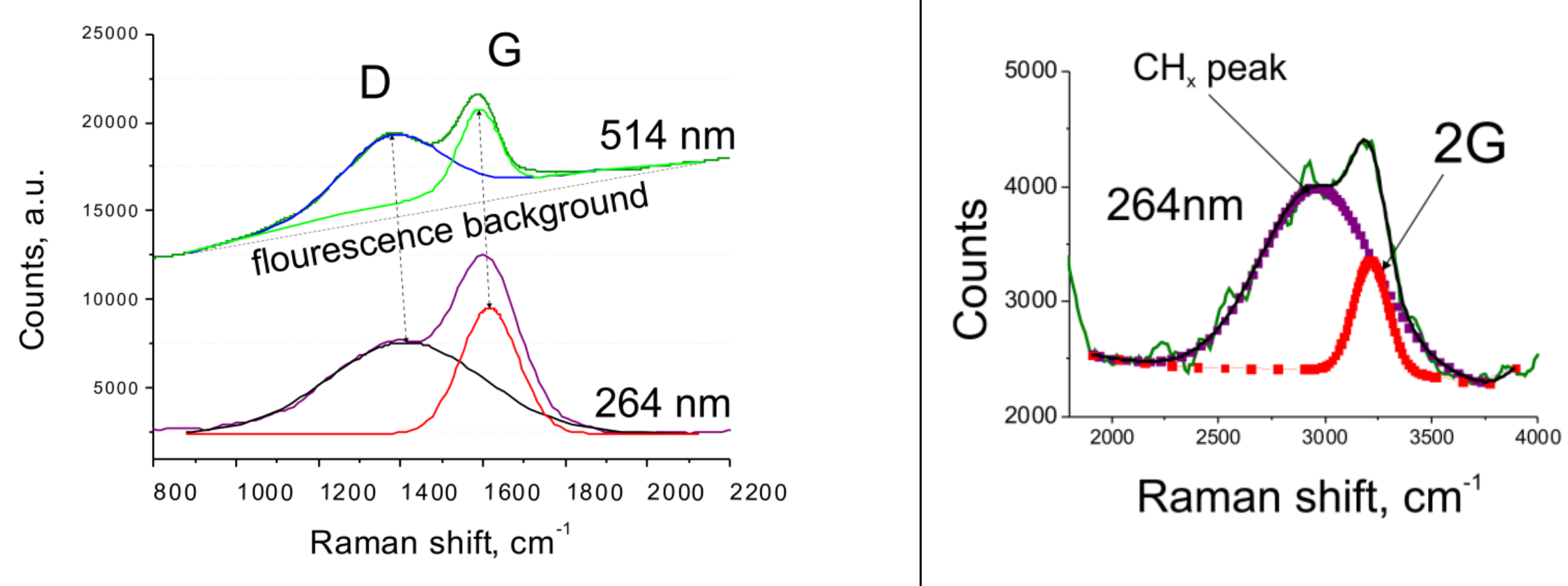

| a | b |
|---|---|

Figure 5. Raman spectra of the carbon film at 514 nm and 264 nm in a range 800-2200 $cm^{-1}$ (a) and in UV-Raman spectra in range 2000-4000 $cm^{-1}$ (b). Peak caused by sp3 $CH_x$ stretching in UV-Raman (b) clearly visible.

*Corresponding author. E-mail: dolgov.adonix@gmail.com; phone: +31622892158 (Netherlands)

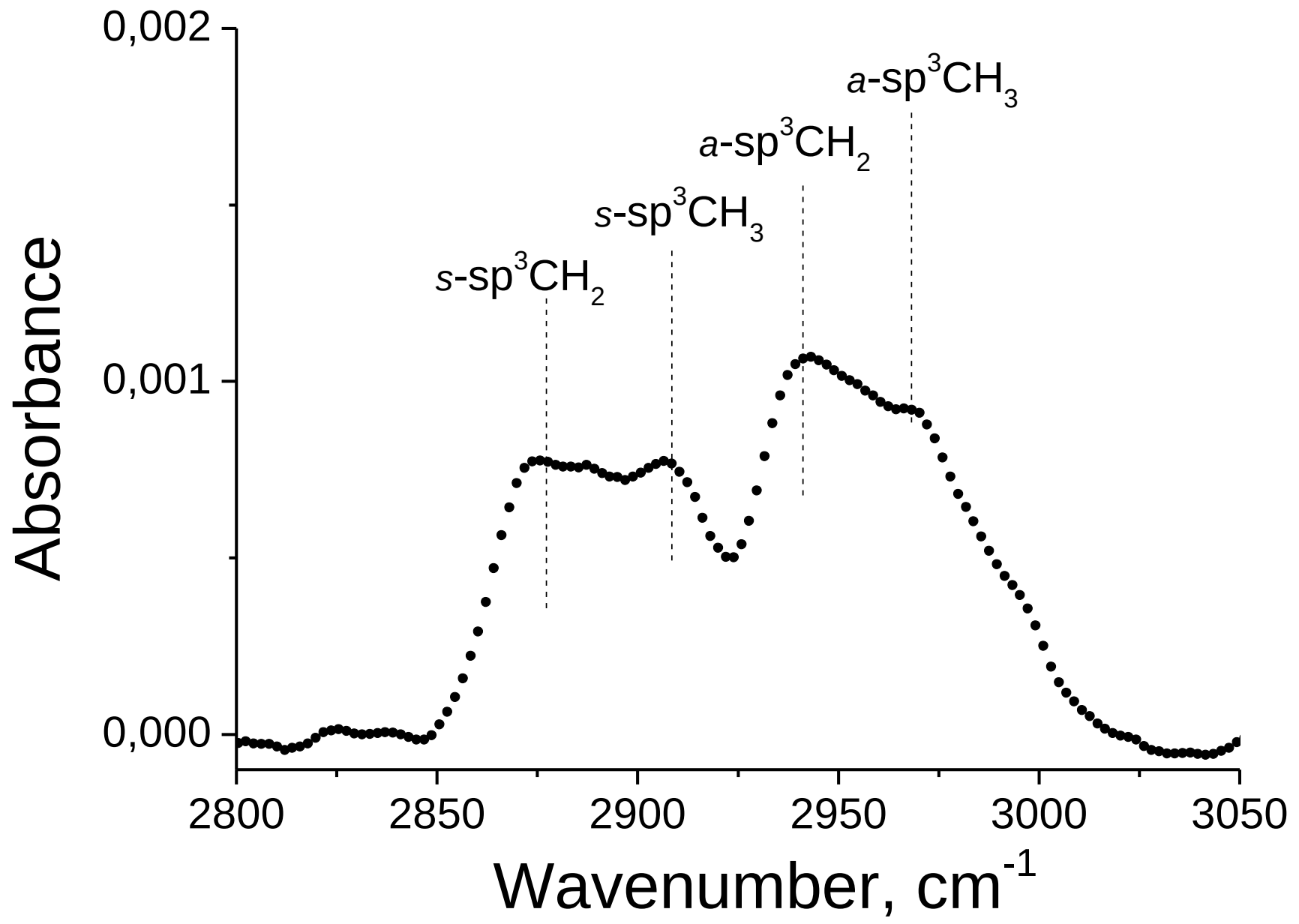

Figure 6. RAIRS absorption spectrum of the collector film in 2800-3000 $cm^{-1}$ range. Spectral resolution is ~4 $cm^{-1}$.

Thus, summarizing, carbon film deposited on mirrors that are plasma-facing (e.g., the collector mirror) will not be polymer-like amorphous carbon. Instead, our results show that the film is more likely to have the properties of hydrogenated diamond-like (DLCH) carbon coatings. These coatings are known to be quite resistant to chemical and physical etching processes, especially compared to standard amorphous carbon.  This result resonates well with studies of thin films (less than 20 nm) reported by prof. Nishiyama's group [20]. It was observed that the films were hydrogenated and deeper layers of EUV-induced carbon contamination demonstrate a diamond-like structure.

The plasma etching was performed to confirm that the carbon film deposited on the collector is hard and inert, like DLCH films. The collector mirrors were cleaned with $H_2$ plasma in an inductively coupled plasma (ICP) reactor. The etch-rate of the collector-grown carbon was compared to the etch rate of magnetron sputtered carbon. The control samples consisted of a Si/Mo multilayer substrate with a ~10 nm thick of amorphous carbon, and a silicon wafer with a ~20 nm thick layer of amorphous carbon. Three samples were simultaneously etched in a plasma reactor, shown in Figure 7.

*Corresponding author. E-mail: dolgov.adonix@gmail.com; phone: +31622892158 (Netherlands)

Ion energy spectra and flux were measured using a DC&RF floating retarded field energy analyzer (RFEA) [21-23]. The plasma density and temperature were measured using an RF compensated Langmuir probe. Through this combination of techniques, the composition, energy spectrum, and flux of ions, incident on the sample surface are accurately known and can be used to characterize of ion cleaning rates and mechanisms in detail.

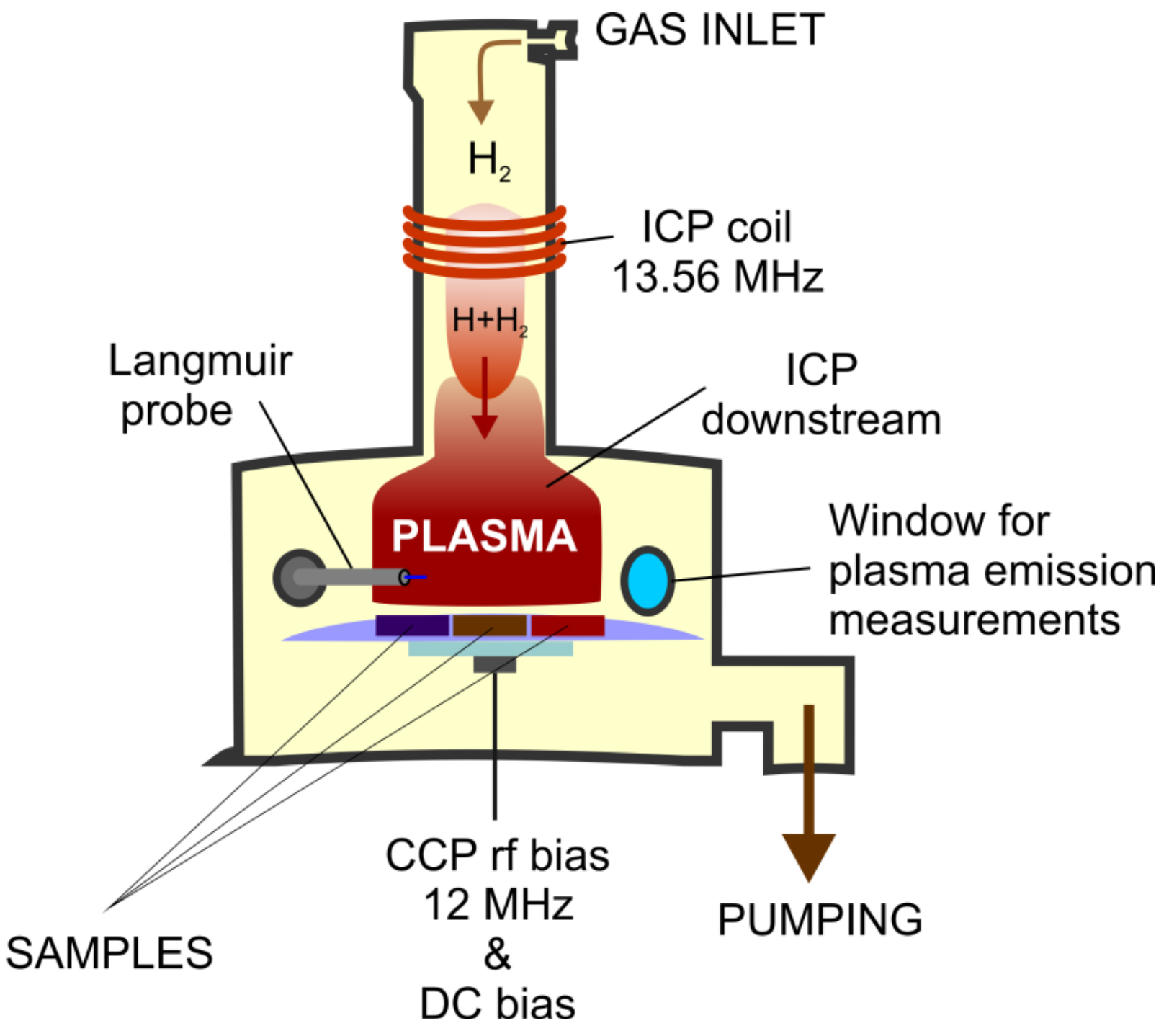

Figure 7. ICP downstream reactor. The plasma excited at 13.56 MHz by 3-turn coil around the quartz tube with an inner diameter of 80 mm. The gas pressure was maintained at 100 mTorr. Samples are placed in ICP discharge downstream at capacitively coupled plasma (CCP) rf-biased electrode. Due to a blocking capacitance and an asymmetric CCP configuration, the electrode can be DC biased and RF (12 MHz) biased.

Our plasma characterization **(see supplementary document)** shows that the sample is exposed to an ion flux that is dominated by $H_3^+$. The composition of ions incident on the samples is presented in Figure 8. The ions have a well peaked energy spectrum, which is established by an external bias field. The full width half maximum of the ion energy spectrum ranges from 4 to 45 eV, depending on the peak ion energy. The ion integral flux was in the range of $10^{13}$-$3{\cdot}10^{14}$ ions/($cm^2{\cdot}s$), depending on the applied bias. The ion energy distribution functions (IEDFs) at the electrode under conditions identical to the conditions that the samples were exposed to are shown in Figure 9. In addition to ions, the surface is also exposed to atomic hydrogen. The flux of atomic hydrogen was found to be $f_H$=8·10$^{18}$ at/(cm$^2$·s).

*Corresponding author. E-mail: dolgov.adonix@gmail.com; phone: +31622892158 (Netherlands)

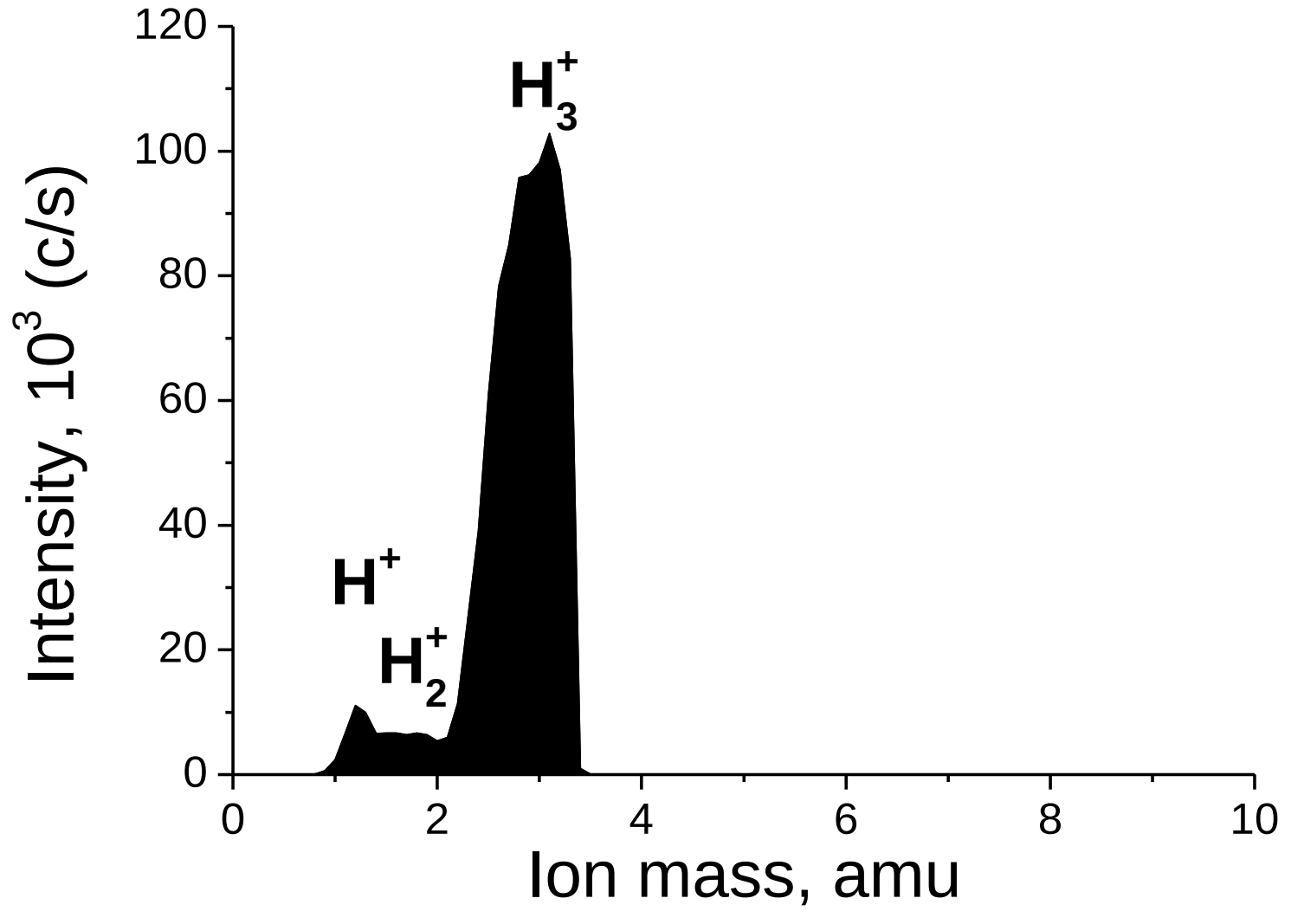

Figure 8. Measured mass-spectrum of ions incident on the sample surface from H2 plasma at an H2 pressure of 100 mTorr and 200 W ICP 13.56 MHz discharge.

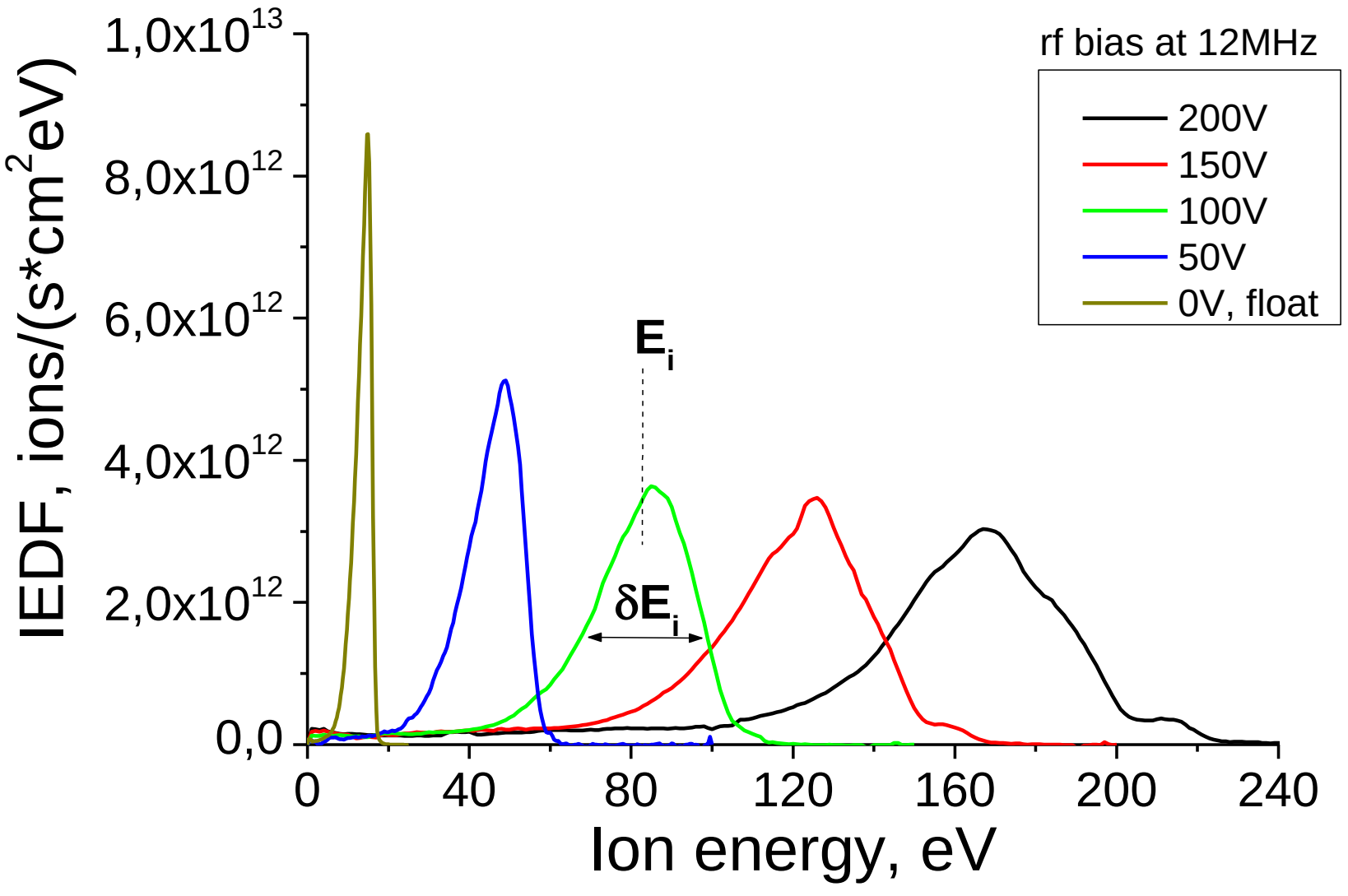

Figure 9. Ion energy distribution function at surface of 12 MHz biased electrode (where samples were placed) at $H_2$ 100 mTorr, 200 W ICP 13.56 MHz discharge and different rf bias voltages.

*Corresponding author. E-mail: dolgov.adonix@gmail.com; phone: +31622892158 (Netherlands)

The sample surface in the plasma is exposed to hydrogen atoms and ions. By measuring the ion dose and etch rate, we estimated the carbon atom removal probability or C yield per incident ion as a function of ion energy (see figure 10). In order to simplify the estimation, the density of all studied carbon films was assumed to be the same at 2 g/cm$^3$. The etch probabilities for magnetron-deposited amorphous carbon (a-C) agree well with the previous data obtained for different $H_2$ plasmas [9]. Therefore, it can be expected that chemical sputtering and reactive ion etching are predominant mechanisms, especially with increasing ion energy [9]. At the lowest energies, the etch rate is constant. This is most likely dominated by atomic hydrogen etching. $H_2$ dissociation mostly occurs in the ICP 13.56 MHz plasma, which was kept constant during the experiment. The estimated C etch probability per H atom was about ~$10^{-6}$ for magnetron-deposited carbon, which is in good agreement with previous studies of plasma cleaning and hot filament H radical cleaning [9,25].

As can be seen in figure 10, the etch probability of EUV collector grown carbon (EUC-C) is notably lower than the magnetron-deposited a-C etch probability. Furthermore, the slopes of the curves in figure 10 (etch probability vs ion energy) are different, indirectly indicating that the carbon in the two films have different binding energies. So the higher energy threshold of chemical sputtering for EUV-C can be explained by a harder carbon structure and, correspondingly, a higher binding energy compared to magnetron sputtered a-C films. For instance, TRIM (TRansport of Ions in Matter) software [26], used for calculating ion interaction with materials (mainly amorphous), assumes 25 eV displacement energy for atoms in stable (hard) carbon material. For amorphous or polymer-like carbons, this energy appears to be notably lower [9]. These experimental results show that the collector film is hard, and together with other data, it is most likely to be a hydrogenated DLC film.

*Corresponding author. E-mail: dolgov.adonix@gmail.com; phone: +31622892158 (Netherlands)

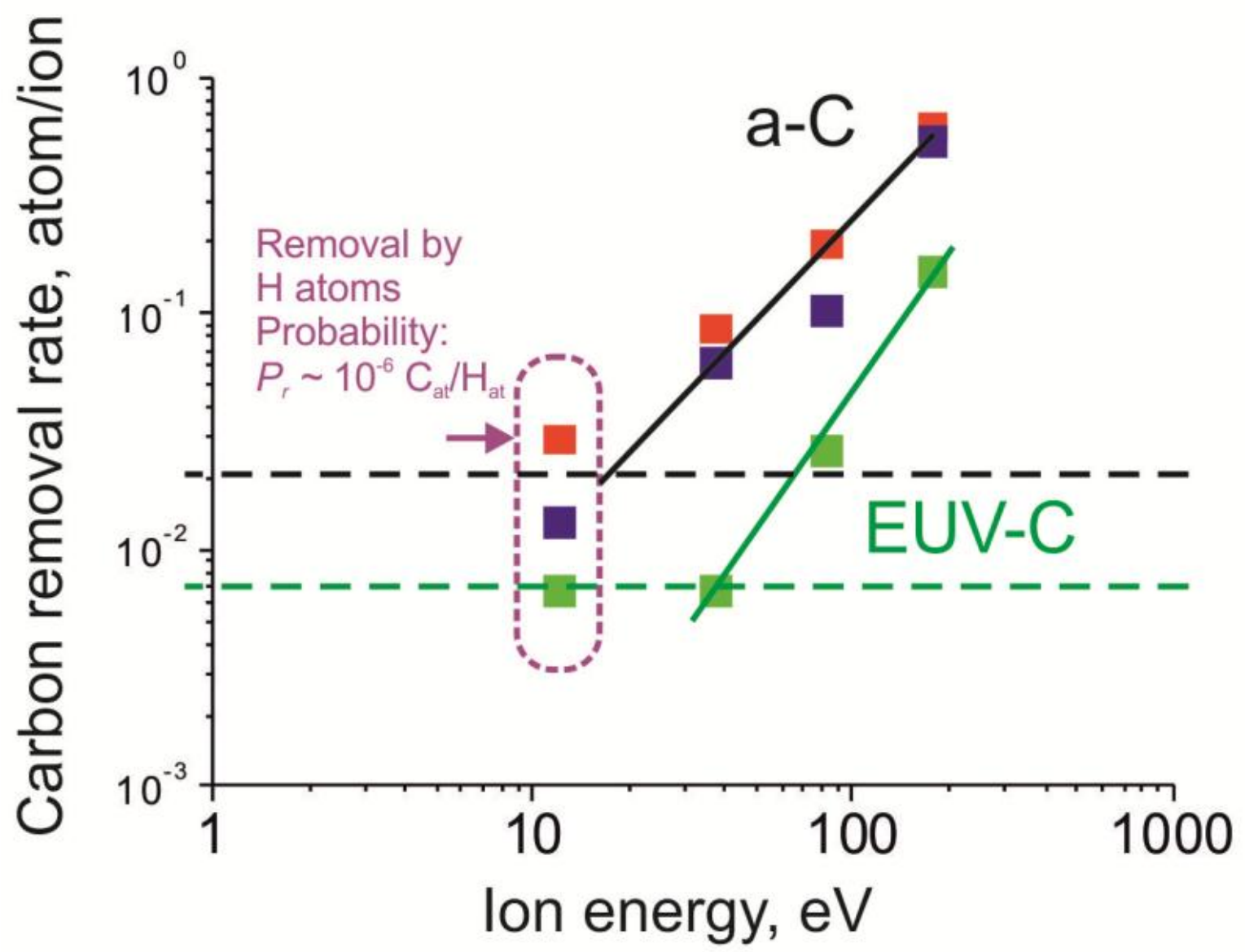

Figure 10. Carbon removal rate (atom/ion) for the different samples as function of ion energy. Dashed lines indicate approximate thresholds of chemical sputtering of amorphous magnetron-deposited and collector films.

To a first approximation the process of collector film formation in an $H_2$ environment can be considered to be similar to the well-known plasma enhanced chemical vapor deposition (PECVD) of hard carbon and DLC coatings. In this deposition process, plasma is used as a source of reactive species, and the substrate and film is subjected to a flux of energetic ions. The similarity between PECVD and EUV-induced deposition can be seen by considering the environmental conditions. The carbon layer is produced due to dissociation of absorbed hydrocarbons by EUV photons and secondary electrons. Dissociation occurs even in high-vacuum conditions ($< 10^{-8}$ Torr) because the residence time of multi-atom hydrocarbon molecules on a surface is long. In our experiments with EUV exposure, $H_2$ ($\sim 10^{-3}$ Torr) flows over the collector mirror, so the deposition proceeds in the presence of $H_2$ and in the presence of a low density hydrogen plasma that is generated by the EUV pulse [9, 27]. The plasma potential, on the other hand, is high: from a few tens of eV up to the photon energy (91.8 eV). Therefore, the energy of incident ions will be close to the plasma potential.

Low-energy hydrogen ions preferentially etch sp2 bound species. This process has been observed in PECVD processing [10], where increasing the ion energy to a few tens of eV often leads to the formation of a hard carbon, enriched in sp3 phase compared to sp2. This occurs

*Corresponding author. E-mail: dolgov.adonix@gmail.com; phone: +31622892158 (Netherlands)

because sp and sp2 carbon reacts with hydrogen radicals and ions more readily than does sp3 bound carbon [28]. In addition to PECVD photo-dissociation of hydrogen molecules under pinch radiation should be taken into account. Simultaneous surface interaction of H atoms with CH3 sp3 bonded radicals leads to increase of sticking coefficient for methyl radicals [29, 30]. These processes lead to enrichment of the film structure by sp3 carbon. Moreover the space in between Z-pinch and the collector, because of the presence of EUV and high energetic Sn ions, acts as the source of hydrogen and hydrocarbon ions and radicals. The 92 eV photon energy is much more than C-C or C-H bond energy. Radiation, hydrogen ions and heavy tin ions interaction can easily cause formation of carbon C+ ions with energy more than 20 eV. This alone is already enough to produce hard sp3 rich carbon coatings. Taking into account water, absorbed on the mirror surface, the film is expected to contain also some C-OH and C(O)OH structures because of photo dissociation and ion induced subplantation [10].

## 4.CONCLUSION

We have studied the carbon film grown from hydrocarbons in an atmosphere of $10^{-3}$ Torr hydrogen during exposure to EUV radiation from a Z-pinch tin discharge. The film contains carbon, oxygen, and hydrogen as the main elements and involves submicron tin droplets. The ~150 nm film is transparent in the visible spectral range and chemically relatively inert in comparison with amorphous magnetron deposited carbon, as shown by its insolubility in nitric acid. Through elemental analysis, Raman spectroscopy, and plasma etching experiments, it has been shown that elements of the film are chemically bound in a mechanically and chemically tough structure, formed by C-C, C-H, C-OH and C(O)OH chemical groups with the significant part of the carbon being in $sp^3$ hybridization. Compared to magnetron deposited carbon, the film also has a higher resistance to chemical and physical sputtering during plasma treatment. The reduced hydrogen plasma removal rate for collector grown films can be up to an order of magnitude lower than for a-C especially for hydrogen ions energies below 100eV. Thus the characteristics of the film obtained by both non-destructive and destructive diagnostics allow us to classify the film as diamond-like coating. One can conclude that carbon contamination of EUV optics that are close to the EUV source and in presence of EUV-induced plasma can be rather hard and are similar to hydrogenated DLC coatings. Such diamond-like amorphous carbon could require ten times more $H_3^+$ ions to remove one C atom during hydrogen plasma treatment. To solve the optics cleaning issue for such type of contamination it is necessary to optimize in-situ cleaning methods, such as those based on using EUV-induced plasma, as considered in [9, 27].

*Corresponding author. E-mail: dolgov.adonix@gmail.com; phone: +31622892158 (Netherlands)

**ACKNOWLEDGMENTS**

This work is part of the research program "Controlling photon and plasma induced processes at EUV optical surfaces (CP3E)" of the "Stichting voor Fundamenteel Onderzoek der Materie (FOM)" which is financially supported by the Nederlandse Organisatie voor Wetenschappelijk Onderzoek (NWO). The CP3E programme is cofinanced by Carl Zeiss SMT GmbH (Oberkochen), ASML (Veldhoven), and the AgentschapNL through the Catrene EXEPT program.

*Corresponding author. E-mail: dolgov.adonix@gmail.com; phone: +31622892158 (Netherlands)

*Corresponding author. E-mail: dolgov.adonix@gmail.com; phone: +31622892158 (Netherlands)

*Corresponding author. E-mail: dolgov.adonix@gmail.com; phone: +31622892158 (Netherlands)